\documentclass[conference,a4paper]{IEEEtran}
\ifCLASSINFOpdf
\else
\fi
\usepackage{algorithmic,algorithm}
\usepackage{latexsym}
\usepackage{url}

\makeatletter
\def\endfor{\end{ALC@for}}
\def\endif{\end{ALC@if}}
\def\endwhile{\end{ALC@while}}
\makeatother
\newcommand{\where}{\mathop{\mathrm{where}}\nolimits}
\newcommand{\powerset}{\mathfrak{P}}
\newcommand{\mapping}{\mathfrak{F}}
\newcommand{\order}{\mathrm{O}}

\def\CA{{\cal A}}
\def\CN{{\cal N}}
\def\CD{{\cal D}}
\def\CS{{\cal S}}

\usepackage{theorem}
\theorembodyfont{\upshape}
\usepackage{amsmath,amssymb}

\newtheorem{definition}{Definition}
\newtheorem{lemma}{Lemma}
\newtheorem{theorem}{Theorem}
\newtheorem{fact}{Fact}
\newtheorem{example}{Example}

\newtheorem{collorary}{Corollary}[theorem]
\usepackage{graphicx}

\def\ed{{\hfill $\diamond$}}
\def\re{\mathtt}
\def\S{Sect. }

\usepackage{color}

\begin{document}
%
\title{Simultaneous Finite Automata:\\ An Efficient Data-Parallel Model\\
for Regular Expression Matching}

\author{\IEEEauthorblockN{Ryoma Sin'ya}
\IEEEauthorblockA{Department of Mathematical\\
and Computing Sciences,
\\Tokyo Institute of Technology \\
Email: shinya.r.aa@m.titech.ac.jp}
\and
\IEEEauthorblockN{Kiminori Matsuzaki}
\IEEEauthorblockA{School of Information,\\
Kochi University of Technology\\
Email: matsuzaki.kiminori@kochi-tech.ac.jp}
\and
\IEEEauthorblockN{Masataka Sassa}
\IEEEauthorblockA{Department of Mathematical\\
and Computing Sciences,
\\Tokyo Institute of Technology\\
Email:  sassa@is.titech.ac.jp}}


%


\maketitle

\begin{abstract}
Automata play important roles in wide area of computing and the growth
of multicores calls for their efficient parallel implementation.  Though
it is known in theory that we can perform the computation of a finite
automaton in parallel by simulating transitions, its implementation has
a large overhead due to the simulation.  In this paper we propose a new
automaton called \emph{simultaneous finite automaton} (SFA) for efficient
parallel computation of an automaton.  The key idea is to extend
an automaton so that it involves the simulation of transitions.  
Since an SFA itself has a good property of parallelism, we can develop
easily a parallel implementation without overheads.  We have
implemented a regular expression matcher based on SFA, and it has
achieved over  10-times speedups on an environment with dual hexa-core CPUs 
in a typical case.

{\bf This paper has been accepted at the following conference:
2013 International Conference on Parallel Processing (ICPP-2013),
 October 1-4, 2013 Ecole Normale Supérieure de Lyon, Lyon, France.}
\end{abstract}




%
\IEEEpeerreviewmaketitle

\section{Introduction}

Automata play important roles in theory and practice in a wide area 
of computing. For example the use of non-deterministic or deterministic automata
is crucial in regular expression matching.
Under the growth of multicores, parallelism becomes more and more important.
In previous studies \cite{Kumar:2006:AAM:1151659.1159952,Brodie:2006:SAH:1150019.1136500}, computations of automata are naively executed in parallel when 
both/either of queries and/or data are multiple,
while a single computation of an automaton is executed in sequential. 
To extract more
parallelism, parallelizing an automaton itself would be important.
It has been known for a long time in theory that we can perform the
computation of a finite-state automaton in
parallel~\cite{LF80,Hillis:1986:DPA:7902.7903}.  The basic idea of the
parallelization is to simulate all the transitions from all
the possible states speculatively.  However, as reported in previous
studies~\cite{Hol09,Luchaup:2009:MRE:1691138.1691158,Luchaup:2011:SPP:2335656.2335677,DBLP:journals/corr/abs-1210-5093},
such a parallel implementation has a large overhead due to the
speculative simulation.

In this paper, we propose a novel approach for parallelizing the
computation of automata.  The key idea is to extend automata so that
they involve the speculative simulation from all the states.  
We develop new automata named \emph{simultaneous finite automata} (SFA in short)
as extensions of finite-state automata where the states in SFA are given as mappings from states to states of the original automata.
The key property of the SFA is that they essentially involve parallelism and thus we can straightforwardly implement the computation of SFA in parallel.
Though such an extension
may increase the size of automata, we can remove the runtime overhead.
It is worth noting that usually automata are considerably smaller than
data and the runtime speedup outstrips the enlargement of automata.

We can systematically construct an SFA from either an NFA or a DFA by a technique 
similar to the so-called subset construction technique.  In general, such a construction may
increase the number of states exponentially.  However, for widely-used regular expressions,
the number of states in SFA is no more than the square of that in the original automata.
We show the effectiveness of SFA with
the experiment results of the SFA-based parallel regular expression
matching.  Our SFA-based implementation has almost no overhead and
achieved over 10-times speedups on an environment with dual hexa-core CPUs
with respect to the DFA-based sequential implementation in a typical case.

The contributions of this paper are summarized as follows.
\begin{itemize}
\item We proposed a new automaton, simultaneous finite automaton, for parallel regular expression matching (\S \ref{sec:sfa}).
      By using SFA, we can compute regular expression matching simply in parallel
      without overheads (Algorithm~\ref{code:pcompsfa}). This SFA-based parallel regular-expression matcher
      is available online~\cite{regen}.
\item We developed an algorithm for constructing SFA from NFA or DFA (Algorithm~\ref{code:corcons}).
      Since the algorithm is a natural extension of the subset construction algorithm, we can apply
      known implementation techniques for it.
\item The only concern of SFA is the size explosion with respect to DFA or NFA.
   We show that almost all the SFA are small enough for practical regular expressions in the SNORT rulesets.
   We also discuss the cases that SFA have as many states as the upper bound.
\end{itemize}

The rest of the paper is organized as follows.
We introduce the basic idea of automata in \S \ref{sec:preliminaries}, and
we review the parallelization method based
on the speculative simulation in \S \ref{sec:priorworks}.
In \S \ref{sec:sfa}, we define the simultaneous finite automata and
discuss their properties. In \S \ref{sec:implementation}, we develop the
implementation of SFA: a construction method and an application to
parallel regular expression matching.
In \S \ref{sec:experiment}, we show the experimental results on SFA's
size, scalability, and overheads.
In \S \ref{sec:discussion}, we discuss the algebraic characterization
of SFA for the theoretical upper bound of the number of states.
Finally, we conclude the paper in \S \ref{sec:conclusion}.

\emph{Remarks on automata theory.}
Automata theory has been deeply studied for a long time and 
there exist many extended models of automata in terms of parallelism.
Some examples are {\it parallel finite automata}~\cite{Stotts94parallelfinite}, {\it concurrent finite automata}~\cite{Jantzen+07b}, and {\it alternating finite automata}~\cite{Chandra:1981:ALT:322234.322243}.
These models are extension for dealing with parallel/concurrent events,
and they are \emph{not} for implementing parallel matching of an automaton.
The SFA in this paper is a new automata for discussing data-parallel regular expression matching.


\section{Preliminaries}\label{sec:preliminaries}
\subsection{Notation}
In this paper, we describe definitions and algorithms with symbols in the basic set theory.
Some things to note: $|A|$ denotes the size of set $A$ (number of their elements).  $\powerset(A)$ is the power set of $A$ and  $|\powerset(A)| = 2^{|A|}$ holds. $\mapping(A, B)$ denotes all the mappings from $A$ to $B$ ($f: A \rightarrow B$) and $|\mapping(A, B)| = |B|^{|A|}$ holds. In particular, $\mapping(A, A)$ is called a \emph{transformation} of $A$, and $\mapping(A, \powerset(A))$  is called a \emph{correspondence} of $A$. 
We define \emph{function composition} $\circ$ on transformations and correspondences as follows:
\begin{eqnarray*}
f, g \in \mapping(A, A), \forall a \in A \;\;\;\;\;\; (f \circ g)(a)
 &:=& f(g(a)),\\
f, g \in \mapping(A, \powerset(A)), \forall a \in A \;\;\;\;\;\; (f \circ g)(a) &:=& \bigcup_{b \in g(a)}f(b).
\end{eqnarray*}
We also define \emph{reverse composition} $\bullet$ as $f \bullet g := g \circ f$.
Here, note that function composition and reverse composition are always associative.

\subsection{Finite Automata}
We briefly introduce some basics of automata theory according to \cite{Sak}.
First we give the definition of nondeterministic and deterministic finite automata.
\begin{definition}\upshape
A {\it nondeterministic finite automaton} (NFA) $\CN$ is a quintuple
$\CN = (Q, \Sigma, \delta, I, F)$, where $Q$ is a finite set of states, $\Sigma$ is a set of input symbols, $\delta$ is a transition function of type $Q \times \Sigma \rightarrow \powerset(Q)$, $I \subseteq Q$ is a set of initial states, and $F \subseteq Q$ is a set of final states. 
\ed
\end{definition}
\begin{definition}\upshape
A {\it deterministic finite automaton} (DFA) $\CD$ is a special case of NFA, where $I$ and every image of $\delta$ are singletons:
\begin{flushleft}
\hfill $\displaystyle
 |I| = 1 \quad
\land \quad
 \forall q \in Q, \forall \sigma \in \Sigma \left[ |\delta(q, \sigma)| = 1 \right].
$ \hfill \hbox{}\ed
\end{flushleft}
\end{definition}

We may say the number of states in an automaton as the \emph{size} of the automaton, and we denote the size of automaton $\CA$ as $|\CA|$.
We introduce $\widehat \delta$ for an extended transition function over input texts:
\begin{eqnarray*}
\widehat \delta(q, \sigma w) &:=& \bigcup_{q' \in \delta(q, \sigma)} \widehat \delta(q', w),\\
\widehat \delta(q, \epsilon) &:=& \{q\}~\mbox{.}
\end{eqnarray*}
The symbol $\epsilon$ denotes empty word and transition over $\epsilon$ does nothing.
We also introduce bound transition function $\delta^\sigma, \widehat\delta^w: Q
\rightarrow \powerset(Q)$ defined by follows:
\begin{eqnarray*}
\delta^\sigma(q) &:=& \delta(q, \sigma),\\
\widehat\delta^w(q) &:=& \widehat\delta(q, w)~\mbox{.}
\end{eqnarray*}

If $p \in \widehat \delta(q, w)$ is a transition of automaton $\cal{A}$, $w$ is said to be the \emph{label} of the transition and we will write $q \xrightarrow[\cal{A}]{w} p$ (or simply $q \xrightarrow{w} p$ if it is unambiguous).
\begin{definition}\upshape
A \emph{computation} $c$ in $\CA$ is a sequence of transitions, which can be written as follows:
\[
 c := q_0 \xrightarrow{\sigma_1} q_1 \xrightarrow{\sigma_2} q_2 \xrightarrow{\sigma_3} \cdots \xrightarrow{\sigma_n} q_n~\mbox{.}
\]
A word in $\Sigma^*$ is \emph{accepted} by $\CA$ if it is the label of a computation that begins at an initial state and ends at a final state in $\CA$.
\ed
\end{definition}
\begin{definition}\upshape
$L(\cal{A})$ denotes the set of all the words accepted by $\cal{A}$:
\begin{flushleft}
\hfill $\displaystyle
L(\CA) = \left\{w ~|~ \exists q \in I, \exists p \in F \left[ q \xrightarrow[\cal{A}]{w} p \right]\right\}.
$\hfill \hbox{}\ed
\end{flushleft}
\end{definition}

We say two automata $\CA$ and $\CA'$ are equivalent if $L(\CA) = L(\CA')$ holds. 
The following theorem shows that there exists an equivalent DFA to every automaton.

\begin{theorem}[Rabin and Scott\cite{RS59}]
Every automaton $\cal{A}$ is equivalent to a DFA $\cal{D}$. If $\cal{A}$ is finite with $n$ states, $\cal{D}$ can be constructed with at most $2^n$ states.

\begin{IEEEproof}
Let $\CA = (Q, \Sigma, \delta, I, F)$ be an automaton.  We consider an automaton $\CD = (Q_d, \Sigma, \delta_d, I_d, F_d)$: $Q_d$ is $\powerset(Q)$; $\delta_d$ is the additive extension of $\delta$
\[
S \in \powerset(Q), \sigma \in \Sigma \;\;\;\;\;\; \delta_d(S, \sigma) := \displaystyle\bigcup_{q \in S}\delta(q, \sigma)~\mbox{;} 
\]
$I_d$ is a singleton of set $\{I\}$; final states are given by $F_d = \{ S \in \powerset(Q) | S \cap F \neq \emptyset \}$.  The automaton $\CD$ is deterministic. Furthermore, it is equivalent to $\CA$ since we have the following series of equivalences:
\begin{eqnarray*}
  w \in L(\CA) &\Leftrightarrow& \displaystyle \exists q \in I \left[
				 \widehat \delta(q, w) \cap F \neq \emptyset \right]\\
 &\Leftrightarrow& \widehat \delta_d(\{I\}, w) \cap F \neq \emptyset\\
    &\Leftrightarrow& \displaystyle
       \widehat \delta_d(I_d, w) \in F_d \Leftrightarrow w \in L(\CD).
\end{eqnarray*}\\[-2\baselineskip]
\end{IEEEproof}
\end{theorem}

\subsection{Subset Construction and Sequential Computation in DFA}


It is often faster to perform the computation with DFA than to do with NFA.
Given an NFA, we can determinize it by the \emph{subset construction} technique shown in Algorithm~\ref{code:subcons}. 
Starting from the set of initial states, we compute the accessible subset of DFA step by step considering only those states obtained by applying the transition function to the states already calculated.

\begin{algorithm}[t]
\caption{Subset construction}
\label{code:subcons}
\begin{algorithmic}[1]
\REQUIRE Automata $\CA = (Q, \Sigma, \delta, I, F)$
\ENSURE DFA $\CD = (Q_d, \Sigma, \delta_d, I_d, F_d)$ is equivalent to $\CA$
\STATE $Q_d \leftarrow \emptyset, Q_{tmp} \leftarrow \{I\}$
\WHILE{$Q_{tmp} \neq \emptyset$}
\STATE choose and remove a set $S$ from $Q_{tmp}$
\STATE $Q_d \leftarrow Q_d \cup \{S\}$
\FORALL{$\sigma \in \Sigma$}
\STATE $S_{next} \leftarrow \bigcup_{q \in S} \delta(q, \sigma)$
\STATE $\delta_d[S, \sigma] \leftarrow S_{next}$
\STATE{\bf if} $S_{next} \notin Q_d$ {\bf then} $Q_{tmp} \leftarrow Q_{tmp} \cup \{S_{next}\}$
\ENDFOR
\ENDWHILE
\STATE $I_d \leftarrow \{I\}$
\STATE $F_d \leftarrow \{S \in Q_d | S \cap F \neq \emptyset \}$
\end{algorithmic}

\end{algorithm}

Sequential implementation of the computation in DFA is straightforward.
Algorithm~\ref{code:scompdfa} shows the sequential program for the computation in DFA,
in which we use a table $\delta_d[q, \sigma]$ for the transition function.
Note that we store only a single state and reuse it during the computation.

\begin{algorithm}[t]
\caption{Sequential computation of DFA}
\label{code:scompdfa}
\begin{algorithmic}[1]
\REQUIRE DFA ${\cal D} = (Q_d, \Sigma, \delta_d, \{q_0\}, F_d)$, and \hfill\break word $w = \sigma_1\sigma_2\cdots\sigma_n$
\ENSURE $q_{final}$ is the destination such that $q_0 \xrightarrow[{\cal D}]{w} q_{final}$
\STATE $q \leftarrow q_0$
\FOR{$i = 1 \to n$}
\STATE $q \leftarrow \delta_d[q, \sigma_i]$
\endfor
\STATE $q_{final} \leftarrow q$
\end{algorithmic}

\end{algorithm}

Let $\CD$ be the DFA, $\Sigma$ be the set of input symbols, and $n$ be the size of input word.  
Then, the sequential computation in DFA takes $\mathrm{O}(n)$ time, and 
the number of elements in the table of the transition function is $\mathrm{O}(|\CD||\Sigma|)$.


\section{Prior Works: Parallel Computation in DFA with Speculative Simulation}\label{sec:priorworks}
It has been known for a long time that the computation in DFA can be
performed in parallel on parallel random access machines (PRAMs)~\cite{LF80,Hillis:1986:DPA:7902.7903}.
The fundamental idea is the speculative simulation of
transitions in which we consider all the states as initial states.
Such simulation of transitions forms a finite-sized mapping (between sets of states) and composition of finite-sized mappings is associative.  
This associativity in the composition of mappings enables us to perform parallel reduction
for the computation of DFA.

Algorithm~\ref{code:pcompdfa} shows a parallel
implementation of the computation of DFA based on speculative simulation~\cite{Hol09,Luchaup:2009:MRE:1691138.1691158,Luchaup:2011:SPP:2335656.2335677,DBLP:journals/corr/abs-1210-5093}.
The following two points are important in this algorithm. 
First, the mappings $T_i[\,]$ are computed on subwords independently in parallel and they contain transitions from all the states.
Secondly, we can reduce the subresults either in parallel with associative binary operator $\bullet$ or in sequential. 
\begin{algorithm}[t]
\caption{Parallel computation of DFA}
\label{code:pcompdfa}
\begin{algorithmic}[1]
\REQUIRE DFA ${\cal D} = (Q_d, \Sigma, \delta_d, \{q_0\}, F_d)$, number of threads $p$,\\
word $w = \sigma_{11}\cdots\sigma_{1m_1}\sigma_{21}\cdots\sigma_{2m_2}\cdots\sigma_{p1}\cdots\sigma_{pm_p}$
\ENSURE $q_{final}$ is destination such that $q_0 \xrightarrow[{\cal D}]{w} q_{final}$
\FORALL{$i \in [1, 2, \ldots, p]$ {\bf parallel}}
\FORALL{$q \in Q_d$}
\STATE $T_i[q] \leftarrow q$
\endfor
\FOR{$j = 1 \to m_i$}
\FORALL{$q \in Q_d$}
\STATE $T_i[q] \leftarrow \delta(T_i[q], \sigma_{ij})$
\endfor
\endfor
\ENDFOR
\STATE // parallel reduction
\hfil // sequential reduction \hfil
\STATE $T \leftarrow T_1 \bullet T_2 \bullet \ldots \bullet T_p$
\hfil $q_{final} \leftarrow q_0$ \;\;\;\;\;\;\;\;\;\;\;\;\;\;\;\;\;\;\;\;
\STATE $q_{final} \leftarrow T[q_0]$
\hfil {\bf for} $i = 1 \rightarrow p$ {\bf do}
\STATE \hfil
\;\;\;\;\;\;\;\;\;\;\;\;\;\;\;\;\;\;\;\;\;\;\;\;\;\;\;\;\;\;\;\; $q_{final} \leftarrow T_i[q_{final}]$
\end{algorithmic}

\end{algorithm}

Let $\CD$ be the DFA, $n$ be the size of input word, $p$ be the number of processors.
The time complexities of Algorithm~\ref{code:pcompdfa} are
$\order(|\CD|n/p + |\CD|\log p)$ when parallel reduction is used or
$\order(|\CD|n/p + p)$ when sequential reduction is used~\cite{Hol09}.
The coefficient $|\CD|$ comes from the speculative simulation of transitions, and it
means that the parallel implementation no longer runs faster than
the sequential implementation when the size of the DFA is large.

\section{Simultaneous Finite Automata}\label{sec:sfa}
The simulation-based parallel computation of DFA has a large overhead linear to the size of DFA.
In this section, we propose a new model of automata that involve the
simulation of transitions in the definition.  The key idea is that we can evaluate the simulation in advance
in the same way as we evaluate the set of transitions during the construction of DFA from NFA.
The proposed model have a good property for data parallel
computation.

\subsection{Formal Definition}

We call the automaton \emph{simultaneous finite automaton} (\emph{SFA}, in short).
A state in SFA corresponds to a mapping from states to sets of states in the normal finite automata.

\begin{definition}\upshape
\label{def:sfa}
Let $\CA = (Q, \Sigma, \delta, I, F)$ be an automaton.
A \emph{simultaneous finite automaton} (SFA) constructed from $\CA$ is a quintuple $(Q_s, \Sigma, \delta_s, I_s, F_s)$:
\begin{itemize}
\item $Q_s \subseteq \mapping(Q, \powerset(Q))$ is a set of mappings;
\item $\Sigma$ is the same set of symbols as $\CA$;
\item $\delta_s$ is the additive extension of $\delta$ in $\CA$ that is defined as
$
f \in Q_s, \sigma \in \Sigma, \;\; \delta_s(f, \sigma) := \{f \bullet \delta^\sigma\}\mbox{;}
$
\item $I_s \subseteq Q_s$ is a singleton of identity mapping $\{f_I\}$ that satisfies $f_I(q) = \{q\}$ for any $q \in Q$;
\item $F_s \subseteq Q_s$ is defined as $F_s = \{ f \in Q_s ~|~ \exists q \in I \break\left[ f(q) \cap F \neq \emptyset \right] \}$. \ed
\end{itemize}
\end{definition}

By definition, SFA are entirely deterministic. 
As described later, SFA can be regarded as DFA with simultaneity.

\begin{theorem}
\label{thm:sfa_equ}
Every automaton $\cal{A}$ is equivalent to an SFA $\cal{S}$. 
If $\cal{A}$ is finite with $n$ states, $\cal{S}$ can be constructed with at most $2^{n^2}$ states. 
In particular, if $\CA$ is deterministic, $\cal{S}$ can be constructed with at most $n^n$ states.
\begin{IEEEproof}
Let the original automaton be $\CA = (Q, \Sigma, \delta, I, F)$, and the SFA constructed from $\CA$ be $\CS = (Q_s, \Sigma, \delta_s, \{f_I\}, F_s)$.

In addition to the fact that $\CS$ is deterministic, $\CS$ is equivalent to $\CA$ since we have the following series of equivalences:
\begin{eqnarray*}
w \in L(\CA) &\Leftrightarrow& \exists q \in I \left[ \widehat \delta(q,
			   w) \cap F \neq \emptyset \right]\\
 &\Leftrightarrow& \exists q \in I \left[ \widehat \delta_s(f_I, w)(q)
   \cap F \neq \emptyset \right]\\
  &\Leftrightarrow& \widehat \delta_s(f_I, w) \in F_s \Leftrightarrow w
   \in L(\CS).
\end{eqnarray*}

The size of the set of mappings is bounded as $|Q_s| \leq |\mapping(Q, \powerset(Q))| = 2^{|Q|^2}$.  
If $\CA$ is deterministic, transition function is one-to-one correspondence and $|Q_s| \leq |\mapping(Q, Q)| = |Q|^{|Q|}$.
\end{IEEEproof}
\end{theorem}

\subsection{Example}
Here we give an example of an SFA, which corresponds to a DFA.  
Notice that, though the states in SFA have meanings of mappings from states to sets of states in corresponding automaton, we need not to mind it when we compute the transitions in SFA.
In other words, we can compute all the transitions in a finite automaton simultaneously by simply computing the transitions in SFA.

\begin{example}\upshape
\label{example:dfa_sfa}

\begin{figure}[t]
\centering\includegraphics[scale=.333]{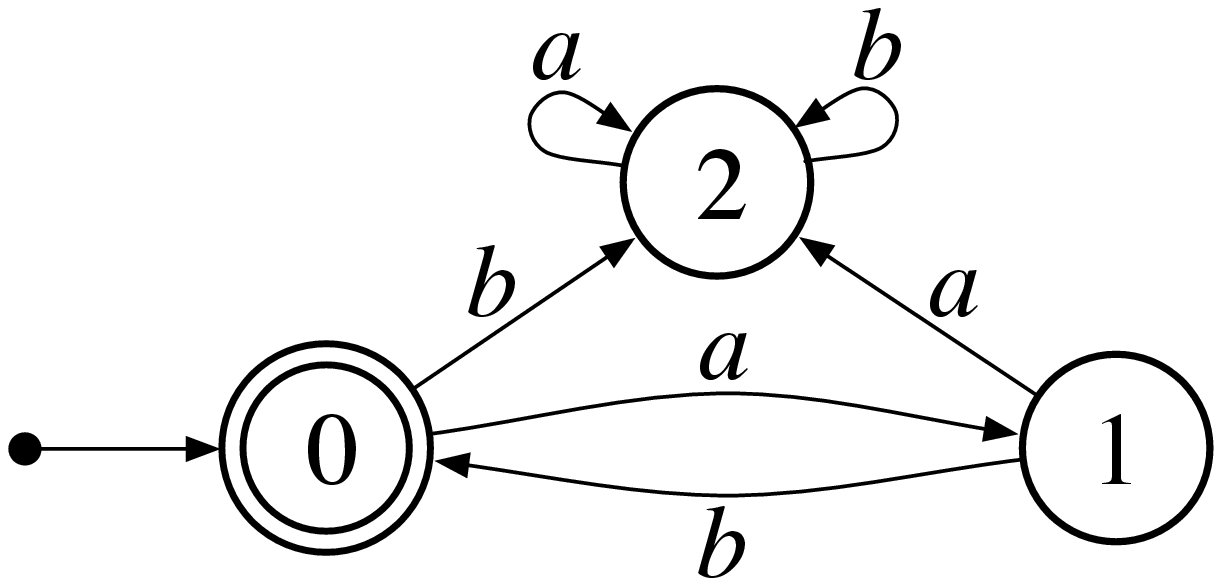}
\caption{$\CD_1: L(\CD_1) = L({\mathtt{(ab)*}})$}
\label{fig:dfa}
\bigskip
\centering\includegraphics[scale=.333]{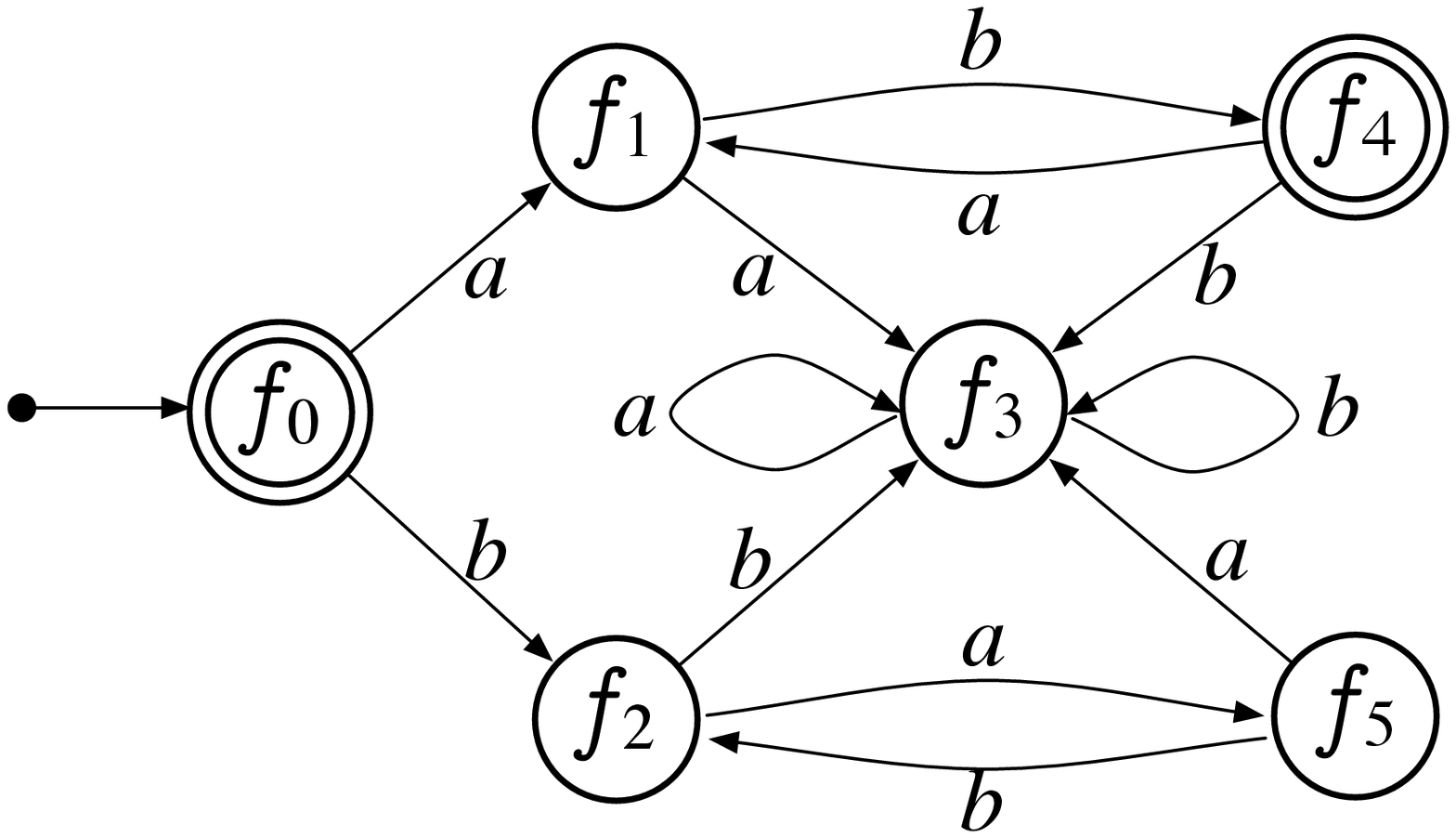}
\caption{$\CS_1: L(\CS_1) = L(\CD_1) = L({\mathtt{(ab)*}})$}
\label{fig:sfa}
\bigskip
\makeatletter\def\@captype{table}\makeatother
\caption{The state mappings of Fig.\ref{fig:sfa}}
\label{table:sfa}
{\normalsize \setlength{\doublerulesep}{.4pt}
\centering\begin{tabular}{@{~}c@{~}|@{~}c@{~}|@{~}c@{~}|@{~}c@{~}|@{~}c@{~}|@{~}c@{~}}
\hline\hline
$f_0$             & $f_1$             & $f_2$             & $f_3$             & $f_4$             & $f_5$\\
\hline
$0 \mapsto \{0\}$ & $0 \mapsto \{1\}$ & $0 \mapsto \{2\}$ & $0 \mapsto \{2\}$ & $0 \mapsto \{0\}$ & $0 \mapsto \{2\}$\\
$1 \mapsto \{1\}$ & $1 \mapsto \{2\}$ & $1 \mapsto \{0\}$ & $1 \mapsto \{2\}$ & $1 \mapsto \{2\}$ & $1 \mapsto \{1\}$\\
$2 \mapsto \{2\}$ & $2 \mapsto \{2\}$ & $2 \mapsto \{2\}$ & $2 \mapsto \{2\}$ & $2 \mapsto \{2\}$ & $2 \mapsto \{2\}$\\
\hline
\end{tabular}
}
\end{figure}

Figure~\ref{fig:dfa} shows DFA $\CD_1$ that accepts $L(\verb|(ab)*|)$.
Figure~\ref{fig:sfa} shows SFA $\CS_1$ equivalent to $\CD_1$ where the states in $\CS_1$ imply the mappings listed in Table~\ref{table:sfa}.
Final states are denoted with doubled circles in these figures. 

Consider the computation of $\CS_1$ over \verb|abab|. By following the states in Fig.~\ref{fig:sfa}, we have transitions $f_0 \xrightarrow[\CS_1]{\mathtt{a}} f_1 \xrightarrow[\CS_1]{\mathtt{b}} f_4 \xrightarrow[\CS_1]{\mathtt{a}} f_1 \xrightarrow[\CS_1]{\mathtt{b}} f_4$.  Here, $f_4(0) = \{0\}$ implies $0 \xrightarrow[\CD_1]{\mathtt{abab}} 0$.  Since the state $0$ is an accepted state in $\CD_1$, $f_4$ is also an accepted state in $\CS_1$.
\ed
\end{example}

\subsection{Data-Parallel Property of SFA}

We finally show an important property of SFA: 
the data-parallel nature in SFA.
For any input text, we can divide it at any points and 
apply the computation of SFA in parallel.

\begin{lemma}\upshape
\label{lemma:divide}
\label{lem1}
Let $\CS$ be an SFA, $f$ be a state in $\CS$, $f_{w_1}$ and $f_{w_2}$
be the states satisfying $f \xrightarrow[\CS]{w_1} f_{w_1}$ and $f_I
\xrightarrow[\CS]{w_2} f_{Iw_2}$. Then the following equation holds:
\begin{eqnarray*}
f \xrightarrow[\CS]{w_1 w_2} f_{w_1 w_2} \Leftrightarrow f_{w_1} \bullet f_{Iw_2} = f_{w_1 w_2}~\mbox{.}
\end{eqnarray*}
\begin{IEEEproof}
By definition, we have
\begin{eqnarray}
 &&f \xrightarrow[\CS]{w_1 w_2} f_{w_1 w_2} ~\Leftrightarrow~ \widehat
 \delta_s(f_{w_1}, w_2) = \{f_{w_1 w_2}\}\label{eq1}\\
 &&\;\;\;\;\;\;\where \; \widehat \delta_s(f, w_1) = \{f_{w_1}\}~\mbox{.}\nonumber
\end{eqnarray}
We can transform the left-hand side as follows by applying the definition of SFA.
\begin{eqnarray}
\widehat \delta_s(f_{w_1}, w_2) &=& 
  \{f_{w_1} \bullet \widehat\delta_s^{w_2} \}
 =
  \{f_{w_1} \bullet (f_I \bullet \widehat\delta_s^{w_2}) \}\nonumber\\
 &=&
  \{f_{w_1} \bullet f_{Iw_2} \}.
\label{eq2}
\end{eqnarray}
We used the fact that $f_I$ is an identity function and the equation
 $\delta_s(f_I, w_2) = \{f_I \bullet \widehat\delta_s^{w_2}\} =
 \{f_{Iw_2}\}$.  
The lemma follows from Equations (\ref{eq1}) and (\ref{eq2}).
\end{IEEEproof}
\end{lemma}

This lemma enables us to introduce the following important theorem about the data-parallelism of the SFA.

\begin{theorem}
The computation in SFA $f_I \xrightarrow[\CS]{w} f$ can be derived by any division of label $w = w_1 w_2 \ldots w_n$.
\begin{IEEEproof}
Computation $f_I \xrightarrow[\CS]{w = w_1 w_2 \cdots w_n} f$ can be decomposed into the following equation by Lemma \ref{lem1}:
\begin{equation*}
f = f_{w_1} \bullet f_{w_2} \bullet \cdots \bullet f_{w_n} \;\;\; \where \;\;\; f_I \xrightarrow[\CS]{w_i} f_{w_i} \;\;\;\;\;\; (i = 1, \ldots, n)~\mbox{.}
\end{equation*}
Each computation $f_I \xrightarrow[\CS]{w_i} f_{w_i}$ has no dependency on the other computations and these composition is associative.  Hence, computation in SFA can be performed in a data-parallel manner. 
We call this method {\it parallel computation} in SFA.
\end{IEEEproof}
\end{theorem}

In the following of the paper, we may classify the SFA in terms of the original automaton.
We call the SFA constructed from NFA as \emph{N-SFA}, and that from DFA
as \emph{D-SFA}.


\section{Implementing SFA}\label{sec:implementation}

\subsection{Construction of SFA from Finite Automaton}

Algorithm~\ref{code:corcons} shows how we can construct an SFA from a finite automaton.
We name the algorithm \emph{correspondence construction} after the subset construction algorithm (Algorithm~\ref{code:subcons}) that constructs a DFA from an NFA.
The correspondence construction algorithm is very similar to the subset construction algorithm,
and the main difference in line 6 of Algorithm~\ref{code:corcons}:
we compute a mapping $f_{next}(q)$ for all the states in the original automaton.
If the original automaton is deterministic, then the image of the
transition function is a singleton and we can simplify the line 6 as follows.
\[
 q \in Q \;\;\;\;\; f_{next}(q) := \delta(q', \sigma) \; \where \; \{q'\} = f(q) \mbox{.}
\]

\begin{algorithm}[t]
\caption{Correspondence construction}
\label{code:corcons}
\begin{algorithmic}[1]
\REQUIRE Automaton $\CA = (Q, \Sigma, \delta, I, F)$
\ENSURE SFA $\CS = (Q_s, \Sigma, \delta_s, I_s, F_s)$ is equivalent to an automaton $\CA$
\STATE $Q_s \leftarrow \emptyset, Q_{tmp} \leftarrow \{f_I\}$
\WHILE{$Q_{tmp} \neq \emptyset$}
\STATE choose and remove a mapping $f$ from $Q_{tmp}$
\STATE $Q_s \leftarrow Q_s \cup \{f\}$
\FORALL{$\sigma \in \Sigma$}
\STATE $q \in Q \;\;\; f_{next}(q) := \bigcup_{q' \in f(q)} \delta(q', \sigma)$
\STATE $\delta_s[f, \sigma] \leftarrow f_{next}$
\STATE {\bf if} $f_{next} \notin Q_s$ {\bf then} $Q_{tmp} \leftarrow Q_{tmp} \cup \{f_{next}\}$
\ENDFOR
\ENDWHILE
\STATE $I_s \leftarrow \{f_I\}$
\STATE $F_s \leftarrow \{f \in Q_s | \exists q \in I | f(q) \cap F \neq \emptyset \}$
\end{algorithmic}

\end{algorithm}

As is the case of the subset construction, the number of the states in the constructed SFA may increase exponentially
compared with that in the original automaton.
As we have stated in Theorem~\ref{thm:sfa_equ}, in the worst case, from an NFA with $n$ states the number of the states in an N-SFA becomes $2^{n^2}$,
and from a DFA with $n$ states the number of the states in a D-SFA becomes $n^n$.
You might consider that these numbers of states dismiss the practical use, but it is not true.
From DFA that correspond to typical regular expressions, fortunately,
the number of states in the constructed D-SFA is no more than the square of that in DFA
(we will show this fact in \S \ref{sec:sizeofsfa}).

The {\it on-the-fly construction} is a well known technique~\cite{Cox1}
in the implementation of an advanced DFA-based matcher. 
The idea of the on-the-fly construction is to construct DFA
during the matching only for the required states, instead of 
constructing full DFA before the matching.
Since on-the-fly construction generates states one by one
after reading symbols, it generates at most $n$ states for input text of length $n$
even if the number of states in DFA explodes.
We can easily apply on-the-fly construction to an SFA-based matcher because
the correspondence construction is a natural extension of the subset
construction.

\subsection{Parallel Computation in SFA}

As we can see from Definition~\ref{def:sfa}, SFA is deterministic in the sense that the image of the transition function is a singleton.
Therefore, we can simply and efficiently implement the computation of SFA by the table-look-up technique.
In addition, from Lemma~\ref{lemma:divide}, we can split the input word at any point and perform the computation of SFA independently in parallel.
After local computation over subtexts, we reduce the results either in parallel with associative binary operator $\bullet$ or in sequential.
Algorithm~\ref{code:pcompsfa} shows the pseudo code of the parallel computation of SFA.

\begin{algorithm}[t]
\caption{Parallel computation of SFA}
\label{code:pcompsfa}
\begin{algorithmic}[1]
\REQUIRE SFA ${\cal S} = (Q_s, \Sigma, \delta_s, \{f_I\}, F_s)$ which is constructed\\
from automaton $\CA = (Q, \Sigma, \delta, I, F)$, number of threads $p$,\\
word $w = \sigma_{11}\cdots\sigma_{1m_1}\sigma_{21}\cdots\sigma_{2m_2}\cdots\sigma_{p1}\cdots\sigma_{pm_p}$
\ENSURE $S_{fin}$ is a set of destinations such that $\forall p \in S_{fin}, \exists q \in I \left[q \xrightarrow[{\CA}]{w} p\right]$ 
\FORALL{$i \in [1, 2, \ldots, p]$ {\bf parallel}}
\STATE $f_i \leftarrow f_I$
\FOR{$j = 1 \to m_i$}
\STATE $f_i \leftarrow \delta[f_i, \sigma_{ij}]$
\endfor
\ENDFOR
\STATE // parallel reduction
\hfil  // sequential reduction \;\;\;\;
\STATE $f_{fin} \leftarrow f_1 \bullet \ldots \bullet f_p$
\hfil \; $S_{fin} \leftarrow I$ \;\;\;\;\;\;\;\;\;\;\;\;\;\;\;\;\;\;\;\;\;\;\;\;\;\;
\STATE $S_{fin} \leftarrow \bigcup_{q \in I} f_{fin}(q)$
\hfil  {\bf for} $i = 1 \rightarrow p$ {\bf do}\;\;\;\;\;\;\;\;\;\;\;\;\;\;\;
\STATE
\hfil \;\;\;\;\;\;\;\;\;\;\;\;\;\;\;\;\;\;\;\;\;\;\;\;\;\;\;\;\;\;\;\;\;\;\;\;\;\;\;\;\; $S_{fin} \leftarrow \bigcup_{p \in S_{fin}} f_i(p)$
\end{algorithmic}

\end{algorithm}

\begin{example}\upshape
\label{example:pcompsfa}
We show how Algorithm~\ref{code:pcompsfa} runs using the SFA $\CS_1$ given in
 Example \ref{example:dfa_sfa}.
Let the number of processors $p$ be 4, and the input word $w$ be ${\mathtt{ababababababab}}$ that is split as $w = w_1 w_2 w_3 w_4$ such that $w_1 = {\mathtt{aba}}$, $w_2 = {\mathtt{baba}}$, $w_3 = {\mathtt{bab}}$, and $w_4 = {\mathtt{abab}}$. In the following, step 1 corresponds to lines 1--5 in Algorithm~\ref{code:pcompsfa} and step 2 corresponds to lines 6--9.
\begin{enumerate}
\item[step 1] For each subword $w_i$, we compute transitions by $\CS_1$ independently in parallel.
For example, on the first processor, we get 
$f_0 \xrightarrow{{\mathtt a}} f_1 \xrightarrow{{\mathtt b}}  f_4 \xrightarrow{{\mathtt a}}  f_1$.  
In the same manner, we get 
$f_0 \xrightarrow{w_2 = {\mathtt{baba}}} f_5$, 
$f_0 \xrightarrow{w_3 = {\mathtt{bab}}}  f_2$, 
and $f_0 \xrightarrow{w_4 = {\mathtt{abab}}}  f_4$.
\item[step 2] We calculate the reduction in parallel on the results of step 1, that is, we calculate $(f_1 \bullet f_5) \bullet (f_2 \bullet f_4)$. 
Here, we can compute the function composition with the mappings in Table~\ref{table:sfa}. 
For example, we get
$(f_1 \bullet f_5)(0) = (f_5 \circ f_1)(0) = f_5(1) = \{1\}$, and similarly, 
$(f_1 \bullet f_5)(1) = \{2\}$ and $(f_1 \bullet f_5)(2) = \{2\}$; as a consequence we get $f_1 \bullet f_5 = f_1$ from these results.
Evaluating the other $\bullet$ operators, we get 
$(f_1 \bullet f_5) \bullet (f_2 \bullet f_4) = f_1 \bullet f_2 = f_4$ as desired. 
\ed
\end{enumerate}
\end{example}

It is worth remarking that in Algorithm~\ref{code:pcompsfa} each thread only deals with a single state
in SFA and just looks up the transition table once for each
character. In Algorithm~\ref{code:pcompsfa}, we have therefore no overhead linear to the number of states
in DFA, which is the defect of Algorithm~\ref{code:pcompdfa}.  The possible overhead is unfortunate cache misses due to the enlargement of the transition table, but the overhead is quite small for practical regular expressions is discussed later.

We can also compute the reduction sequentially: starting from the initial state in the original automaton, we simply compute the states by picking up the states from the mappings obtained in step 1.
In the case of Example~\ref{example:pcompsfa}, we have $(f_4 \circ f_2 \circ f_5 \circ f_1)(0) = (f_4 \circ f_2 \circ f_5)(1) = \cdots = \{0\}$.  We can compute this sequential reduction in $\order(p)$ time, which is independent from the number of states in SFA.

\begin{table*}[t]
\caption{Comparison of complexity}\label{table:complexity}
{\normalsize\setlength{\doublerulesep}{.4pt}
\begin{center}
\begin{tabular}{c | l | p{25em}}
\hline\hline
 Model		 &  State complexity	                & \hfill Computation
 time complexity \hfill~\\
\hline
NFA $\CN$	 &   $|\CN| = \order(m)$	            &   $\order(|\CN|n)$                         \hfill (\cite{ASU} p.165)\\
DFA $\CD$	 &   $|\CD| = \order(2^{|\CN|})$		&   $\order(n)$                       \hfill (Algorithm \ref{code:scompdfa}) \\
             &                                      &   $\order(|\CD|n/p + |\CD|\log p)$  \hfill  (Algorithm \ref{code:pcompdfa}) \\
             &                                      &   $\order(|\CD|n/p + p)$            \hfill  (sequential reduction)\\
N-SFA $\CS_n$ &   $ |\CS_n| = \order(2^{|\CN|^2})$	&   $\order(n/p + |\CN|^3\log{p})$\\
             &                                      &   $\order(n/p + |\CN|p)$ \hfill (sequential reduction)\\
D-SFA $\CS_d$ &   $ |\CS_d| = \order(|\CD|^{|\CD|})$	&   $\order(n/p + |\CD|\log{p})$\\
             &                                      &   $\order(n/p + p)$ \hfill (sequential reduction)\\
\hline
\end{tabular} \\
$m$ is length of regular expression, $n$ is length of input word, $p$ is
 number of threads
\end{center}
}
\end{table*}

Table~\ref{table:complexity} lists the maximum number of states and the execution time.
The last four lines in the table differ in terms of the cost of the reduction.  In parallel reduction for N-SFA, the computation of $\bullet$ operator corresponds to the logical matrix multiplication ($\order(|\CN|^3)$).
In sequential reduction for N-SFA, we evaluate the function one by one, which corresponds to sequential computation of NFA ($\order(|\CN|)$).  In parallel reduction for D-SFA, we need to simulate the transitions for all the states in DFA, and it means we need $\order(|\CD|)$ time for each computation of $\bullet$.
The sequential reduction for D-SFA is the same as the transition of DFA ($\order(1)$).


\section{Experimental Results}\label{sec:experiment}
We have implemented an SFA-based parallel regular expression matcher~\cite{regen}.
It runs in the following four steps:
first it converts a regular expression into an NFA by McNaughton and
Yamada's algorithm~\cite{MN60};
secondly into a DFA by the subset construction (Algorithm~\ref{code:subcons});
thirdly into a SFA by the correspondence construction (Algorithm~\ref{code:corcons});
finally it executes Algorithm~\ref{code:pcompsfa} (with the sequential reduction) specialized to the constructed SFA.

In the following, we show experiment results conducted to confirm the good scalability and small overhead of parallel computation of SFA.
The experiment environment is a PC with two Intel Xeon E5645 CPUs (2.40 GHz,
6 physical cores, SpeedStep/TurboBoost off) and 12 GB DDR3-SDRAM (1333 MHz).
We used CentOS release 5.5 for OS and pthread for the thread library.
In the following results, the throughput and the execution time are of
computation of DFA or SFA, and exclude construction of automata.

\subsection{The size of SFA}\label{sec:sizeofsfa}

The first question that may concern the reader the most would be
``\emph{How large SFA are compared with original DFA for practical regular expressions}?''
To answer this question, we have constructed SFA and DFA 
for over 20000 regular expressions included in the rulesets of 
SNORT network intrusion prevention and detection
system\footnote{http://snort.org/}~\cite{Roes99}, and compared the sizes of automata.

\begin{figure}[t]
\centering\includegraphics[scale=.34,angle=270]{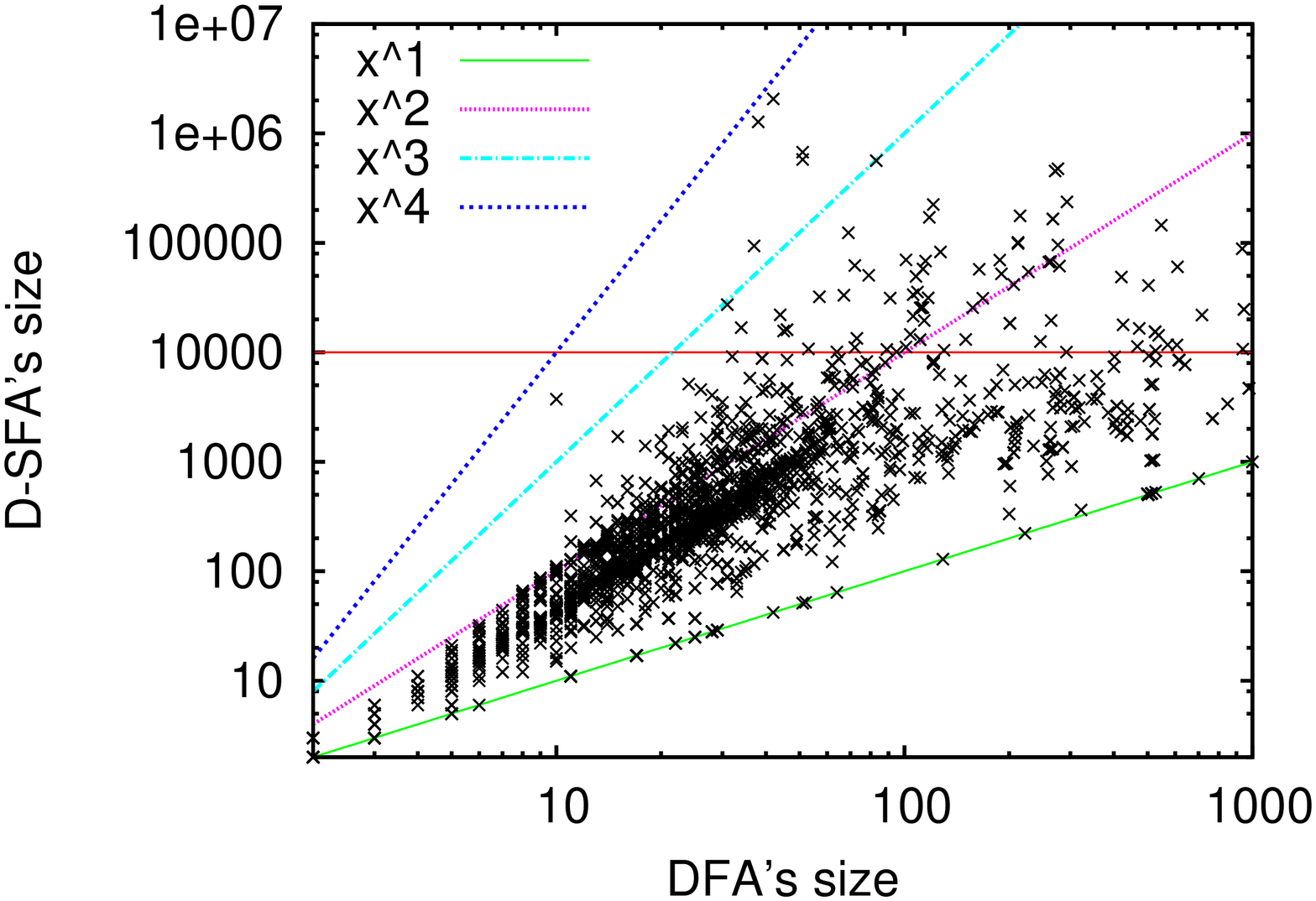}
\caption{The distribution of the size of the minimal DFA and D-SFA on
 SNORT rulesets.}
\label{fig:statenum}
\end{figure}

The details of the experiments are as follows.
The version of the rulesets we used was ``snortrules-snapshot-2940 (03 Feb, 2013)''.
We extracted about 24000 regular expressions from the rulesets, and 
used 20312 regular expressions for the experiments. 
(We did not used too large expressions for which DFA has more than 1000 states,
nor extended expressions that include back references \emph{etc.})
For each regular expression, we constructed a minimized DFA
and then a D-SFA by Algorithm~4.
Figure~\ref{fig:statenum} plots the sizes of D-SFA to the sizes of minimized DFA.

We would like to discuss the number of states in D-SFA from two viewpoints:
absolute size of D-SFA and relative size of D-SFA compared with DFA.
Firstly, only 102 (0.5\%) regular expressions lead to D-SFA that have more than 10000 states.
As we discuss later, current CPUs efficiently compute automata with 10000 states.
Therefore, for almost all the practical regular expressions, we can use D-SFA for efficient parallel matching.

Secondly, for almost all the regular expressions, the number of states in the D-SFA 
is not more than the square of the number of states in the minimal DFA.
Only 279 (1.4\%) regular expressions lead to a D-SFA of over-square size ($|\CS_d| > |D|^2$),
and just 6 regular expressions lead to a D-SFA of over-cubed size ($|\CS_d| > |D|^3$).
These 6 regular expressions have a pattern similar to:
\[
\re{.*(T.*T.*Y.*P.*P.*R.*O.*M.*P.*T.*)}
\] 
in which several \verb|.*| appear in sequence.
For the above regular expression, the size of the minimal DFA is 10 but the size of D-SFA is 3739.
It is worth noting that \emph{no} regular expressions in the rulesets lead to a D-SFA of over-quadruplicate size ($|\CS_d| > |D|^4$).

In theory, the size of a D-SFA $|\CS_d|$ is bounded by $|\CD|^{|\CD|}$
where $|\CD|$ is the size of the DFA from which the D-SFA is constructed
(Table~\ref{table:complexity}).
From the experiment results, however, we conclude that the size of D-SFA 
never grows up exponentially for practical regular expressions.
Of course, in a theoretical perspective, there exist regular expressions that lead to N-SFA or D-SFA
of near upper-bound sizes.
We will discuss them in \S\ref{sec:explosion}.

\subsection{Scalability}

\begin{figure}[t]
\centering\includegraphics[scale=.30]{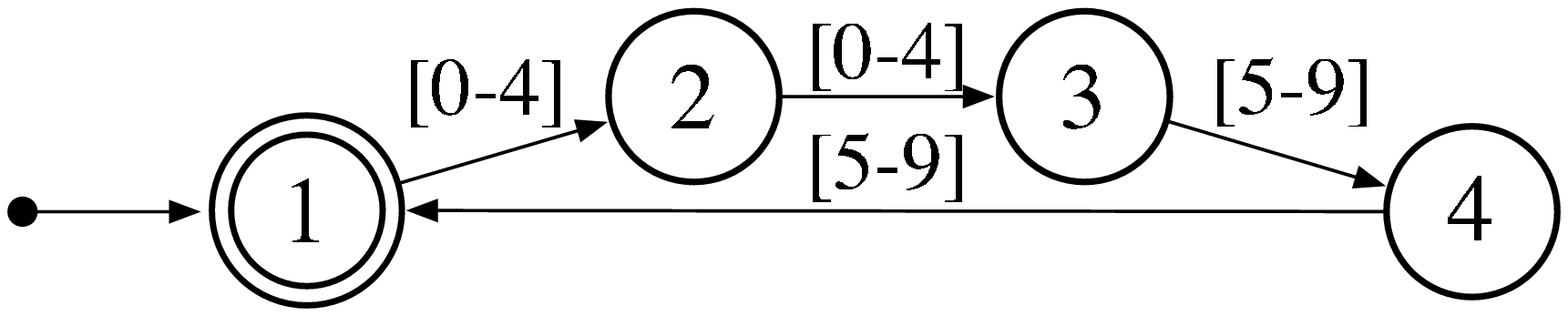}
\caption{The DFA of the regular expression $r_2 = $ \texttt{([0-4]\{2\}[5-9]\{2\})*}}
\label{fig:loop_dfa}
\end{figure}

\begin{figure}[t]
\centering\includegraphics[scale=.30]{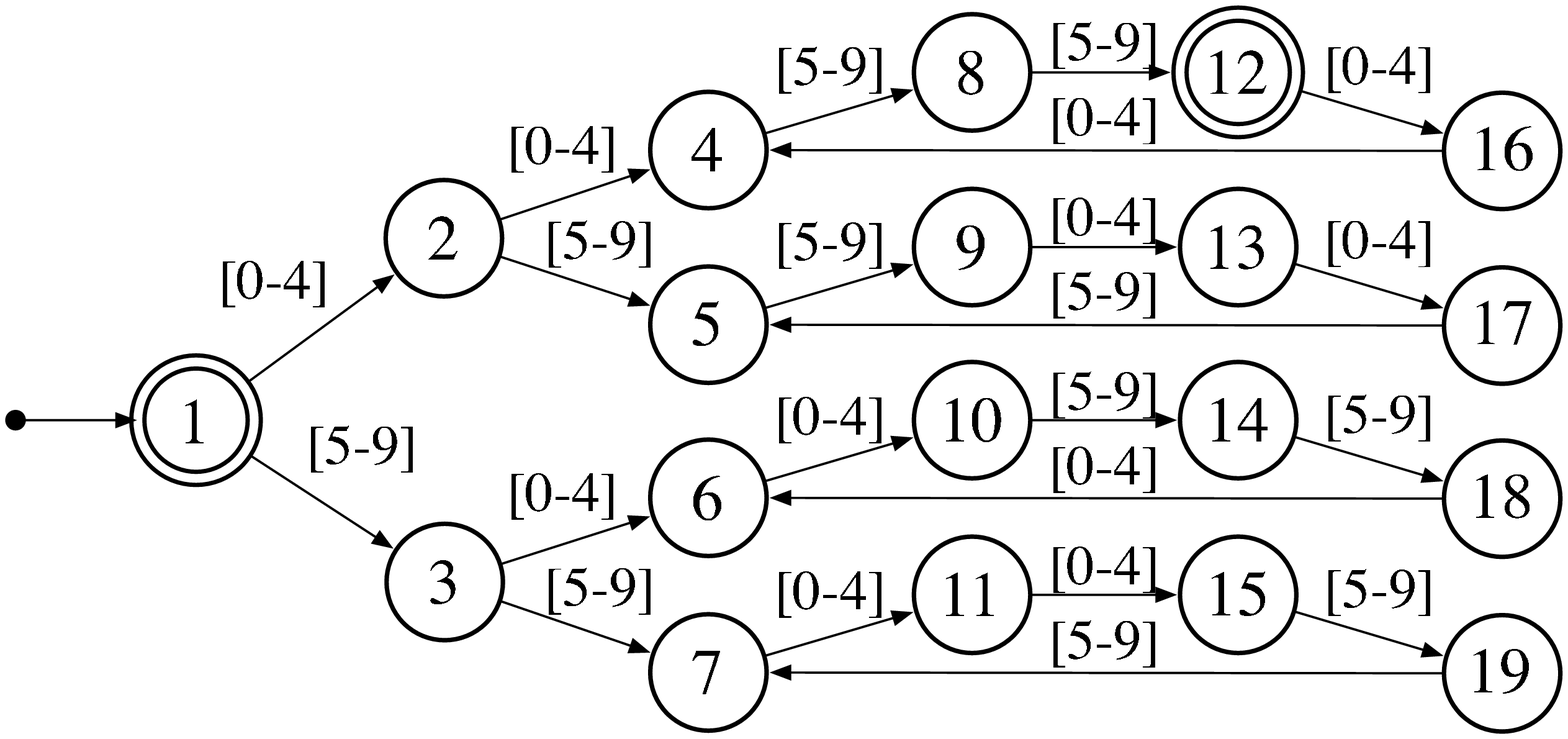}
\caption{The D-SFA of the regular expression $r_2 = $ \texttt{([0-4]\{2\}[5-9]\{2\})*}}
\label{fig:loop_sfa}
\end{figure}

Second question is ``\emph{Does the SFA-based parallel matching scale}?''
We confirmed the scalability of the parallel computation of SFA
with regular expressions in the following form:
\[
 r_n = \re{([0-4]\{}n\re{\}[5-9]\{}n\re{\})*}
\] 
for $n = 5$, $50$, and $500$.  
It is worth noting that the sizes of D-SFA for these expressions 
are almost the square of those of DFA.  For better understanding,
we illustrate the minimal DFA in Fig.~\ref{fig:loop_dfa} and the corresponding
D-SFA in Fig.~\ref{fig:loop_sfa} for the case $n=2$.
The DFA has $2n$ states in a single loop, but the D-SFA
has $2n$ \emph{loops} to distinguish from which state (in DFA) we start.
This is a typical case when we have square-sized D-SFA.

\begin{figure}[tbp]
\centering\includegraphics[scale=.28,angle=270]{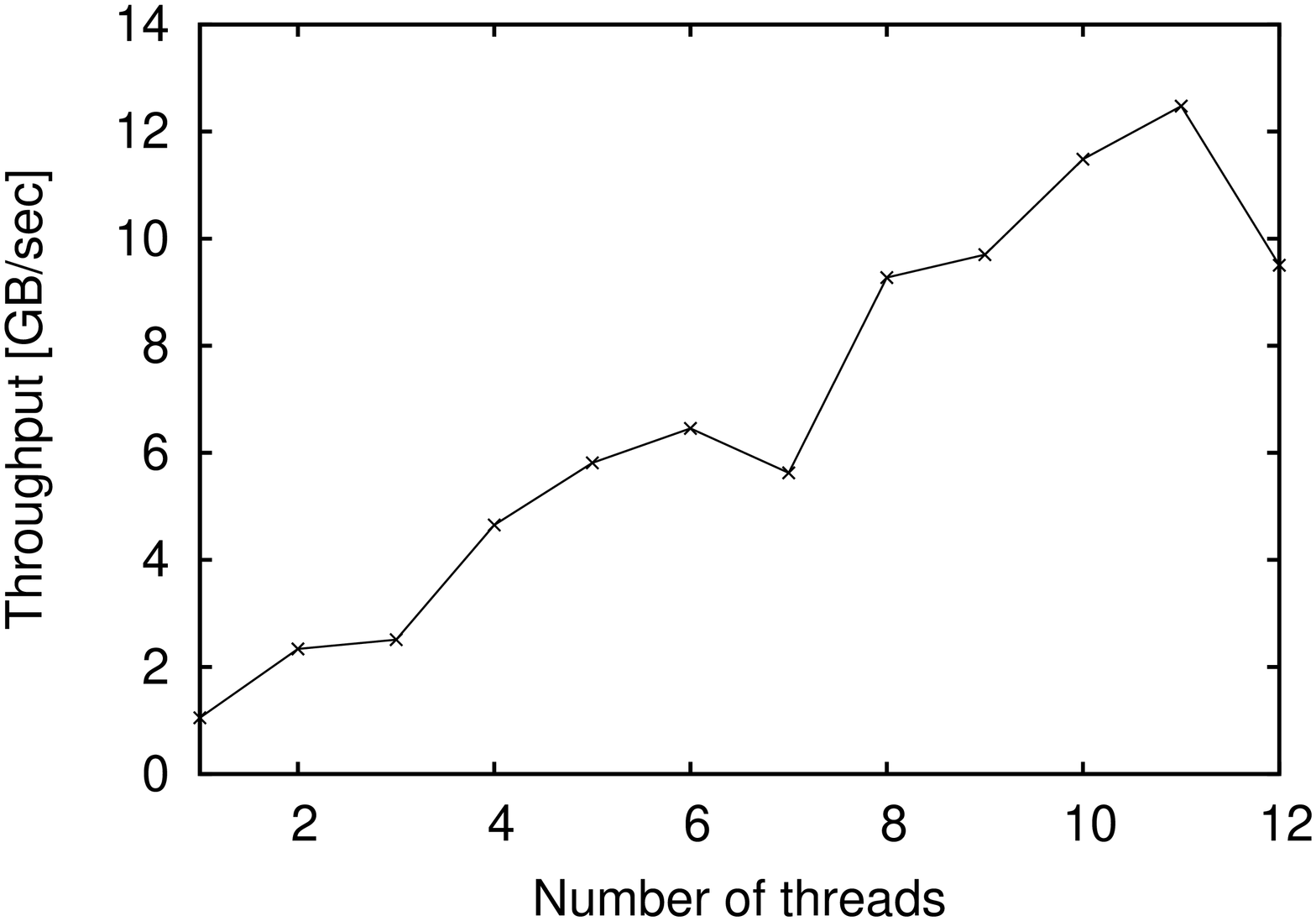}
\caption{$r_5 = $\texttt{([0-4]\{5\}[5-9]\{5\})*}, $|\CD| = 10, |\CS_d| = 109$}
\label{fig:bench5}
\vfill
\centering\includegraphics[scale=.28,angle=270]{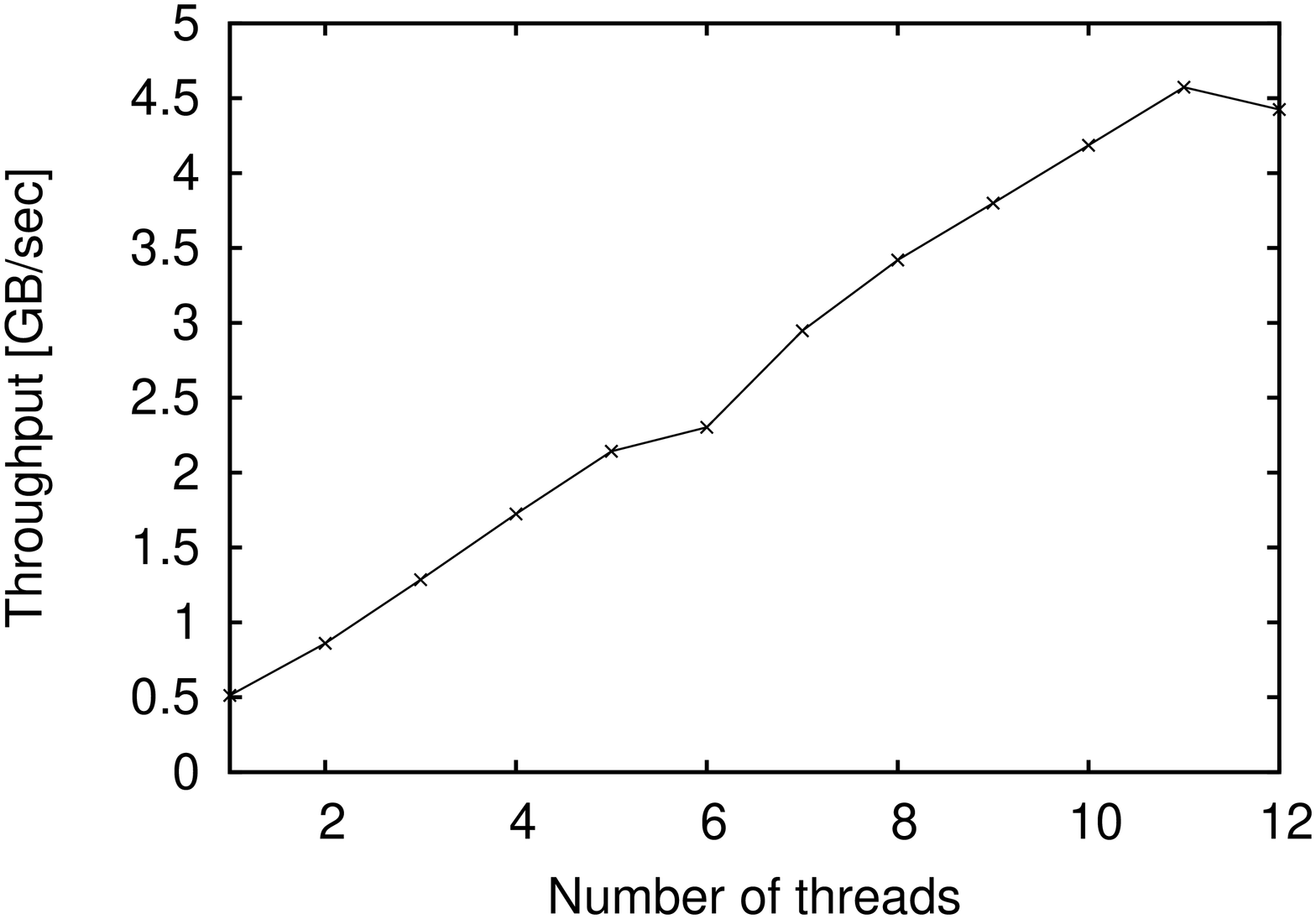}
\caption{$r_{50} = $\texttt{([0-4]\{50\}[5-9]\{50\})*}, $|\CD| = 100, |\CS_d| = 10099$}
\label{fig:bench50}
\vfill
\centering\includegraphics[scale=.28,angle=270]{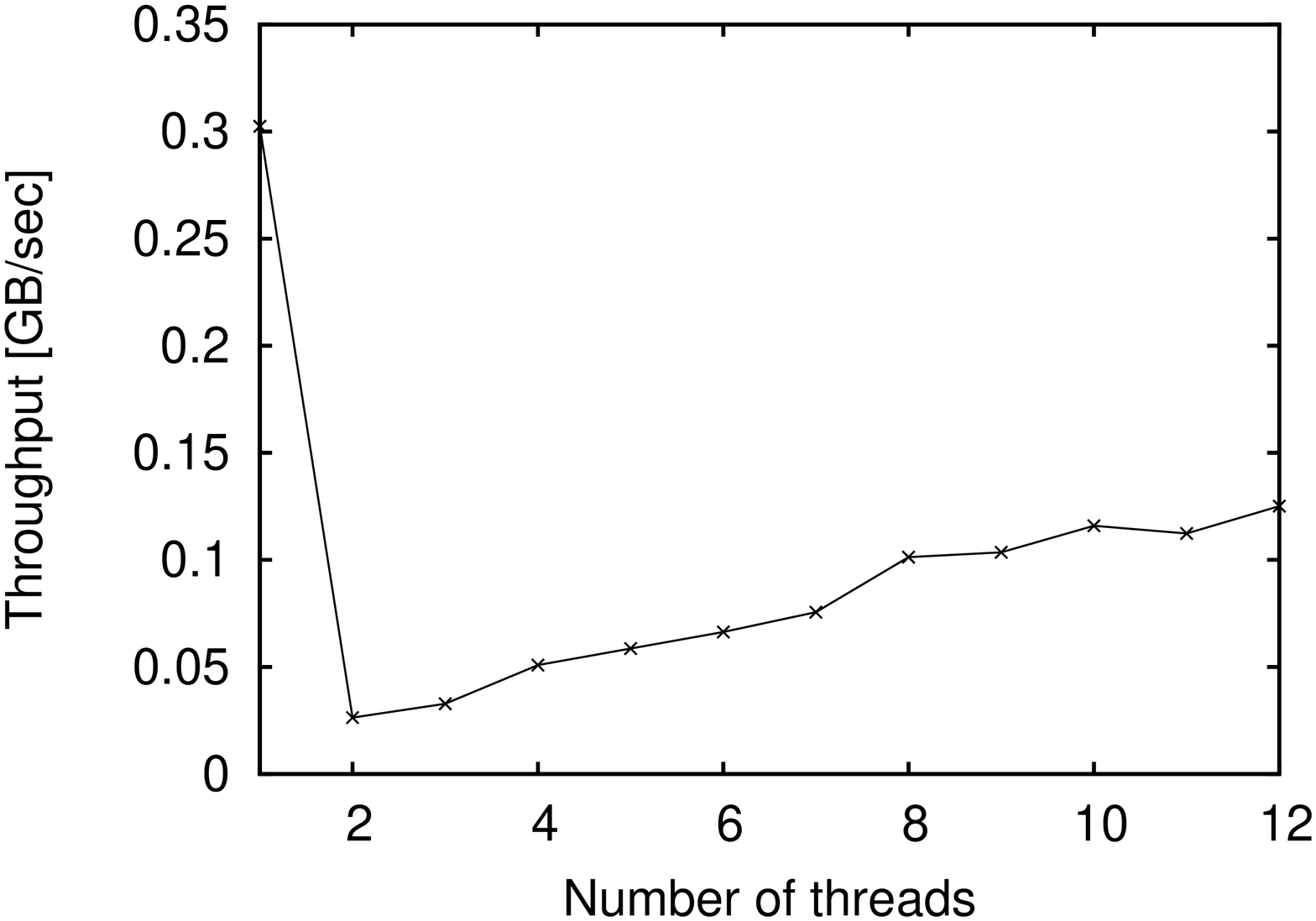}
\caption{$r_{500} = $\texttt{([0-4]\{500\}[5-9]\{500\})*}, $|\CD| =
 1000, |\CS_d| = 1000999$}
\label{fig:bench500}
\vfill
\centering\includegraphics[scale=.28,angle=270]{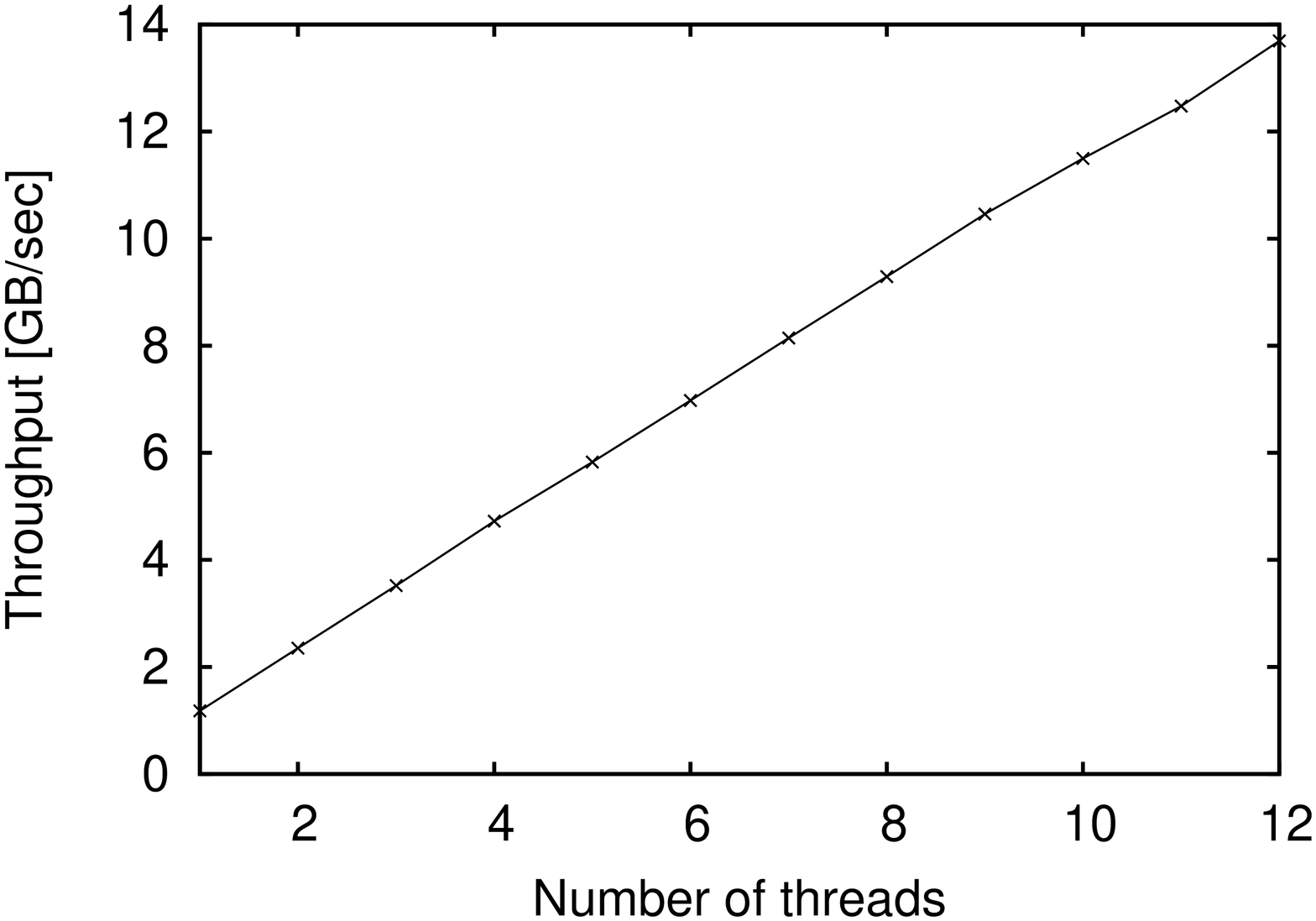}
\caption{$r_a = $\texttt{([0-4]\{500\}[5-9]\{500\})*|a*}, $|\CD| = 1002,
 |\CS_d| = 1001000$, input text is the repetition of ``a'' (1GB)}
\label{fig:bencha}
\end{figure}

Figures~\ref{fig:bench5} to \ref{fig:bencha} show the throughput of the DFA or D-SFA. 
Note that the results with one thread were of DFA (and not D-SFA).
The input texts were 1GB string accepted by those automata, and
every character was read exactly once.
The input texts were stored on the memory before the execution.


As seen in Figures~\ref{fig:bench5} and \ref{fig:bench50}, the SFA-based
parallel matching scales well up to 12 threads (with respect to the sequential
DFA-base matching).
However, in Fig.~\ref{fig:bench500}, the SFA-based parallel matching
ran slower (even with 12 threads) than sequential DFA-based matching.
The difference between them was the size of SFA (and DFA). 
For $r=50$, the number of states in SFA was 10099 and parallel matching
performed well for this size.
For $r=500$, the number of states in SFA was 1000999 while the number of states in DFA was 1000.
In our implementation, the transition table occupied 1KB for each state (256 symbols times 4 bytes).
For $r=500$ the transition table for SFA was 1GB and thus it overflowed the CPU cache (The L3 cache of the CPU was 12MB).

It is worth noting that the large size of SFA does not always mean the poor performance.
It is often the case that transitions are done among small number of states,
and then we can avoid cache misses fortunately.
Figure~\ref{fig:bencha} shows the experiment results for the regular expression
\verb$([0-4]{500}[5-9]{500})*|a*$ and input text being a repetition of ``\verb|a|''.
Although the number of states in SFA was the biggest (1001000),
it achieved the best throughput.
In this case, the transitions were done in a single state and cache misses were avoided.

\subsection{Overheads}
\begin{figure}[t]
\centering\includegraphics[scale=.3,angle=270]{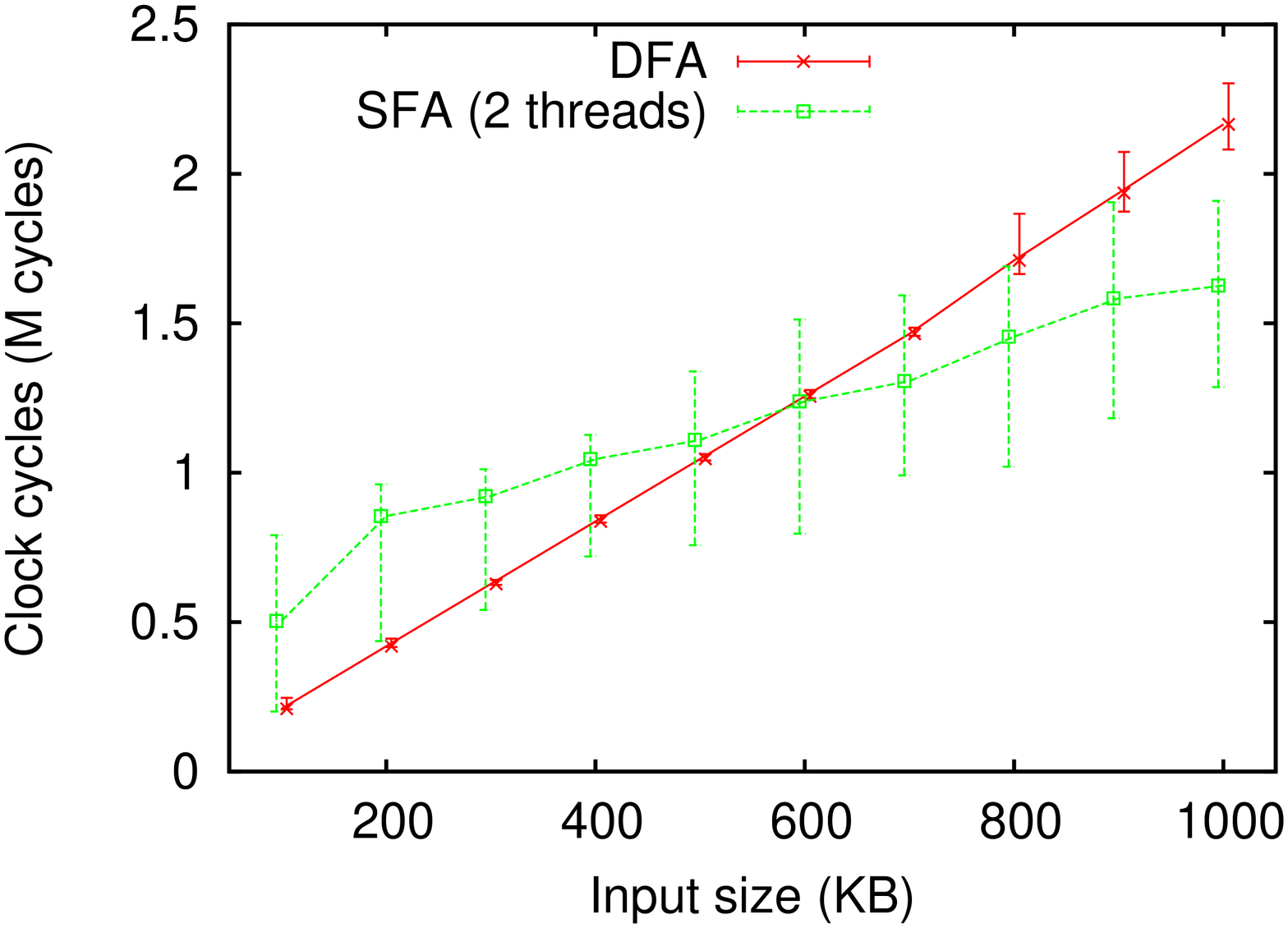}
\caption{Execution times on small inputs}
\label{fig:small}
\end{figure}

We conducted another set of experiments using a smaller input to evaluate the overhead.
Figure~\ref{fig:small} shows the execution times of the sequential computation of DFA and the parallel computation of SFA with two threads.
The execution times of the parallel computation includes the creation of threads and the reduction.
Here we used regular expression \verb|(([02468][13579]){5})*| (the size of DFA is 10, and the size of SFA is 21).
Though the execution time of the parallel computation swings caused by interfere between threads, but the parallel computation runs faster in average over 600KB, and completely over 800KB.

Finally we briefly remark on the cost of constructing SFA. 
Table~\ref{table:construction} shows the time required to constructing DFA and SFA
for the regular expressions $ r_n = \re{([0-4]\{}n\re{\}[5-9]\{}n\re{\})*} $.
Though the correspondence construction of D-SFA from DFA is 
slower than construction of DFA because we need to calculate the mapping between states,
it is fast enough to generate about 50000 states per second.
As we have seen in Fig.~\ref{fig:statenum}, D-SFA for almost all the practical regular expressions
are smaller than 10000 states, and thus we can construct them in less than 0.2 seconds. 

\begin{table}[t]
\caption{Times (in sec) for constructing DFA and D-SFA
for $ r_n = \re{([0-4]\{}n\re{\}[5-9]\{}n\re{\})*} $}
\label{table:construction}
{\normalsize\setlength{\doublerulesep}{.4pt}
\begin{center}\begin{tabular}{ c | r | r | r }
\hline\hline
 & $r_5$ & $r_{50}$ & $r_{500}$ \\\hline
 DFA~$\CD$ & 0.0003 & 0.0019 & 0.0187 \\
 $|\CD|$ & 10 & 100 & 1000 \\\hline
 D-SFA~$\CS_d$ & 0.0020 & 0.2020 & 23.937 \\
 $|\CS_d|$ & ~~~~109 & ~~10099 & 1000999 \\
\hline
\end{tabular}\end{center}
}
\end{table}

\section{Discussion}\label{sec:discussion}
\subsection{Syntactic monoid}
In this paper, we proposed simultaneous finite automaton (SFA) as a data-parallel model of regular expression matching. 
SFA are natural extensions of finite automata on the automata theory. 
In addition, SFA can be regarded as special cases of DFA that include the structure of a syntactic monoid~\cite{Pin,Sak}, which is an algebraic characterization of the regular language.
We would like to emphasize that SFA will bridge the gap between the practice of automata and abstract theory of syntactic monoid.

The size of a syntactic monoid for a regular language is called \emph{syntactic complexity}. 
Indeed, syntactic complexity of a regular language is also the size of
a minimal SFA of the identical language. 
So far, syntactic complexity has received less attention than state complexity that is the size of a minimal DFA~\cite{Brz11}.

As we have shown in this paper, SFA provide a data-parallel model of regular expression matching, and thus we can say that syntactic complexity is also \emph{parallel complexity} of regular expressions. 
We expect that syntactic complexity gets more attentions for establishing the theory over automata and their parallelization.

\subsection{The state explosion problem: an algebraic approach}\label{sec:explosion}
\begin{figure}[t]
\centering\includegraphics[scale=.40]{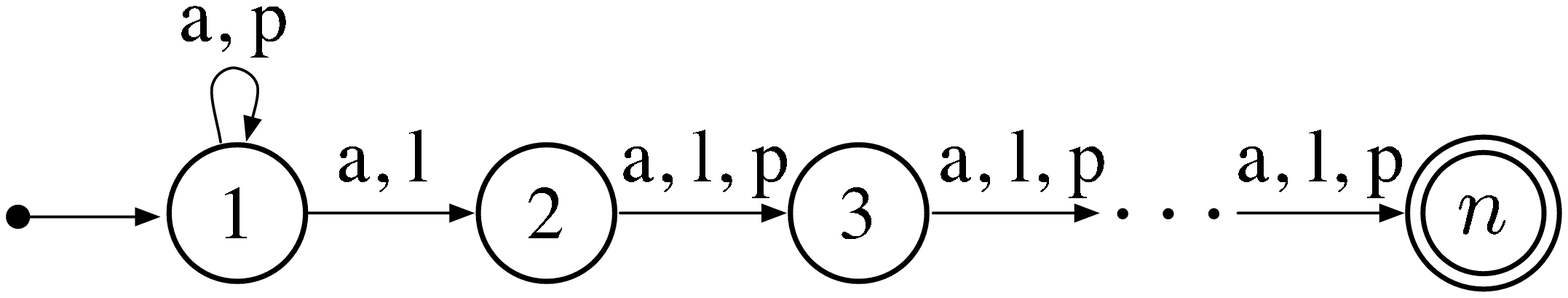}
\caption{The NFA $\CN_{ex \ref{ex:upper_dfa}}$ of the regular expression $e =
 \re{[ap]*[al][alp]\{}n-2\re{\}}$}
\label{fig:upper_dfa}
\end{figure}
\begin{figure}[t]
\centering\includegraphics[scale=.40]{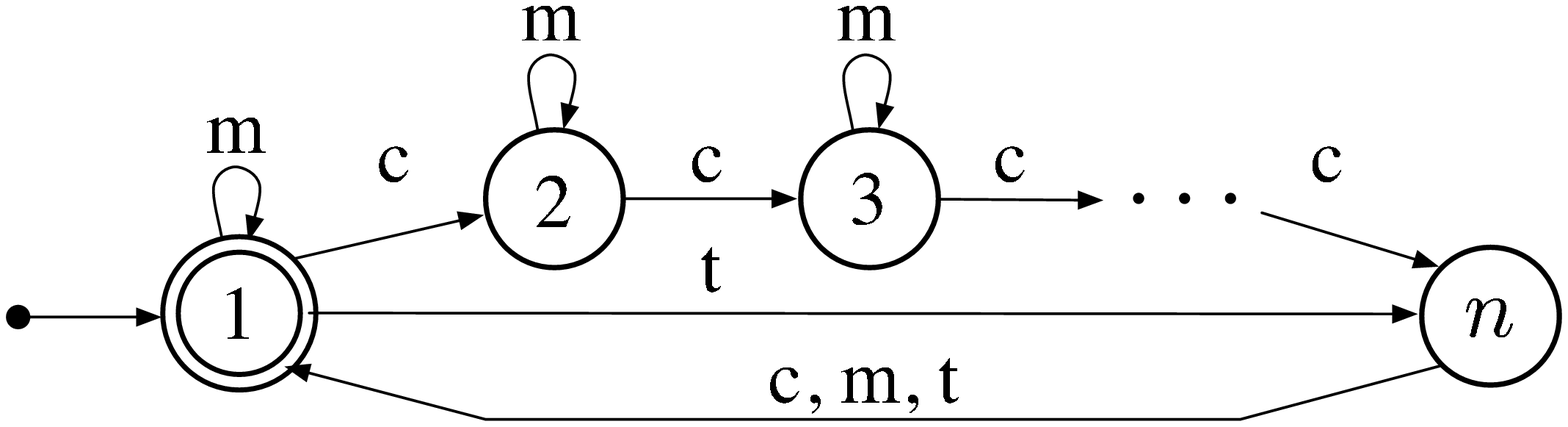}
\caption{The minimal DFA $\CD_{ex \ref{ex:upper_dsfa}}$ of the regular expression $e =
 \re{(m|(t|c([mt]*c)\{}n-2\re{\})[cmt])*}$}
\label{fig:upper_dsfa}
\end{figure}

Here we discuss the theoretical upper bound of the number of states in SFA.
First, we see an example in which we construct a DFA from an NFA followed by
a D-SFA from the DFA.

\begin{example}\label{ex:upper_dfa}
Consider $\Sigma = \{\re{a,l,p}\}$ and the regular expression $e =
 \re{[ap]*[al][alp]\{}n-2\re{\}}$. Figure~\ref{fig:upper_dfa} shows the
 NFA $\CN_{ex \ref{ex:upper_dfa}}$ of the regular expression $e$.

Let us represent the set of states in NFA by a bit-sequence of length $n$.
Then, the initial set of states in Fig.~\ref{fig:upper_dfa} is $1\underbrace{00 \cdots 0}_{n-1}$.
The symbols $\re{a}$ and $\re{l}$ make the following transitions from the initial set of states:
\begin{eqnarray*}
 1\underbrace{00 \ldots 0}_{n-1} &\xrightarrow{\re{a}}& 11\underbrace{00 \ldots 0}_{n-2}~\hbox{, and}\\
 1\underbrace{00 \ldots 0}_{n-1} &\xrightarrow{\re{l}}& 01\underbrace{00 \ldots 0}_{n-2}~,
\end{eqnarray*}
and the symbol $\re{p}$ makes the following transitions:
\begin{eqnarray*}
 11\underbrace{00 \ldots 0}_{n-2} &\xrightarrow{\re{p}}& 101\underbrace{00 \ldots 0}_{n-3}~\hbox{, and}\\
 01\underbrace{00 \ldots 0}_{n-2} &\xrightarrow{\re{p}}& 001\underbrace{00 \ldots 0}_{n-3}~.
\end{eqnarray*}
Notice that the symbols $\re{a}$ and $\re{l}$ correspond to arithmetic shift and logical shift
and the symbol $\re{p}$ corresponds to partial shift applied to bit-sequences from second bit.
With these three shift operations, we can generate all the bit-sequences of 
length $n$ from the initial sequence.
 Hence the minimal DFA
 $\CD_{ex \ref{ex:upper_dfa}}$ of $\CN_{ex \ref{ex:upper_dfa}}$
 satisfies $|\CD_{ex \ref{ex:upper_dfa}}| = 2^{|\CN_{ex
 \ref{ex:upper_dfa}}|}$. 
\ed


\end{example}
By Example \ref{ex:upper_dfa}, we obtain the following fact.
\begin{fact}\label{fact:upper_dfa}
If $|\Sigma| \geq 3$, then there exists a regular expression $e$ over $\Sigma$ whose
 NFA $\CN$ and minimal DFA $\CN$ satisfies $|\CD| = 2^{|\CN|}$.\ed
\end{fact}

Based on a similar idea, we can find a regular expression
for which a D-SFA has as many states as the theoretical upper bound
from the size of DFA.

\begin{example}\label{ex:upper_dsfa}
Consider $\Sigma = \{\re{c,m,t}\}$ and the regular expression $e =
 \re{(m|(t|c([mt]*c)\{}n-2\re{\})[cmt])*}$. Figure~\ref{fig:upper_dsfa}
 shows the minimal DFA $\CD_{ex \ref{ex:upper_dsfa}}$ of the regular
 expression $e$. The minimal D-SFA $\CD_{ex \ref{ex:upper_dsfa}}$ of
 $\CS_{ex \ref{ex:upper_dsfa}}$ satisfies $|\CS_{ex \ref{ex:upper_dsfa}}| =
 |\CD_{ex \ref{ex:upper_dsfa}}|^{|\CD_{ex \ref{ex:upper_dsfa}}|}$. \ed
\end{example}
By Example \ref{ex:upper_dsfa}, we obtain the following fact.
\begin{fact}\label{fact:upper_dsfa}
If $|\Sigma| \geq 3$, then there exists a regular expression $e$ over $\Sigma$ whose minimal DFA $\CD$ and minimal D-SFA $\CS_d$ satisfies $|\CS_d| =
 |\CD|^{|\CD|}$.\ed
\end{fact}

Facts \ref{fact:upper_dfa} and \ref{fact:upper_dsfa} mean
the existence of regular expressions with three symbols
that lead to state explosion in the construction of DFA or D-SFA.
Here, we have another question: \emph{Is there a regular expression 
with a constant number of symbols that lead to state explosion 
in the construction of N-SFA from NFA?}
The following fact on the semigroup theory gives a negative
answer to this question.


\begin{fact}[Devadze~\cite{DBLP:bibsonomy_devadze68,IOPORT.06018260}]\label{fact:devadze}
The size of a minimal generating set of the semigroup of $n \times n$ boolean matrices grows exponentially with $n$.\ed
\end{fact}
This fact was first presented by Devadze in 1968, and he described
minimal sets of generators of the semigroup of $n \times n$ boolean
matrices without a proof. Its was proved very recently by Konieczny in
2011~\cite{IOPORT.06018260}.


We stated in the previous section that the states in SFA correspond to elements in syntactic monoid.
Since the syntactic monoid can be represented with boolean matrices and their multiplication
\footnote{See \cite{HK03,Pin} for the relation between the syntactic monoid and boolean matrices.
Theorem~3 in \cite{HK03} is a proof for Fact \ref{fact:upper_dsfa}.
In the semigroup theory, the problem corresponding to Fact \ref{fact:upper_dsfa} is one of basic propositions (\cite{citeulike:2759286}, Exercise 6).}, 
the theorem also applies to the syntactic monoid.

The following fact follows from Devadze's theory.
\begin{collorary}\label{col:upper_nsfa}
To denote a regular expression that leads to an N-SFA $\CS_n$ with $|\CS_n| = 2^{k^2}$ states,
we require an exponential number of states with respect to $k$.\ed
\end{collorary}
Corollary \ref{col:upper_nsfa} means that it is unrealistic 
to find a large regular expression that leads to state explosion in
the construction of N-SFA.



\section{Conclusion}\label{sec:conclusion}
We have defined a novel class of automata called simultaneous finite
automata, and developed an implementation of them for efficient data-parallel
regular expression matching. 
The parallel computation of SFA runs in $\order(n/p + p)$ time or in
$\order(n/p + |\CD|\log p)$ time where $|\CD|$ is the number of states
in DFA, $n$ is the length of input word and $p$ is the number of
threads. 

We tackled SFA's size issue in \S \ref{sec:sizeofsfa}, made experiments
in real world regular expressions (SNORT rulesets), and show that SFA's
size is fully practical in typical case.
We also made experiments with the SFA-based regular expression matcher, and
confirmed good scalability by a factor of over 10 on an environments with dual hexa-core CPUs
and small overhead such that execution with two threads outperforms for input data over 600KB.

Our implementation of the SFA-based parallel regular expression matcher
is available as an open-source software~\cite{regen}, hence anyone can
verify the experimental results in \S \ref{sec:experiment}.

\subsection*{Acknowledgments}
We would like to thank Kazuhiro Inaba for the helpful discussion with
him on the prior works described in \S \ref{sec:priorworks} and
\S \ref{sec:explosion}.

This work was partly supported by the joint project between ANR (France)
and JST (Japan) (project PaPDAS ANR-2010-INTB-0205-02 and JST 10102704).




\end{document}